\documentclass[fleqn,usenatbib]{mnras}

\usepackage{newtxtext,newtxmath}

\usepackage[T1]{fontenc}

\DeclareRobustCommand{\VAN}[3]{#2}
\let\VANthebibliography\thebibliography
\def\thebibliography{\DeclareRobustCommand{\VAN}[3]{##3}\VANthebibliography}

\usepackage{graphicx}	
\usepackage{amsmath}	

\def\msun{\hbox{M$_\odot$}}

\title[UV-dim stars in LMC star clusters]{On the origin of UV-dim stars: a population of rapidly rotating shell stars?}

\author[Martocchia et al.]{
S. Martocchia,$^{1}$\thanks{E-mail: silvia.martocchia@uni-heidelberg.de}
N. Bastian,$^{2,3,4}$
S. Saracino,$^{4}$
S. Kamann$^{4}$
\\
$^{1}$Astronomisches Rechen-Institut, Zentrum f\"ur Astronomie der Universit\"at Heidelberg, M\"onchhofstraße 12-14, D-69120 Heidelberg, Germany\\
$^{2}$Donostia International Physics Center (DIPC), Paseo Manuel de Lardizabal, 4, 20018, Donostia-San Sebasti\'an, Guipuzkoa, Spain\\
$^{3}$
IKERBASQUE, Basque Foundation for Science, 48013, Bilbao, Spain\\
$^{4}$Astrophysics Research Institute, Liverpool John Moores University, IC2 Liverpool Science Park, 146 Brownlow Hill, Liverpool L3 5RF, UK
}

\date{Accepted XXX. Received YYY; in original form ZZZ}

\pubyear{2023}

\begin{document}
\label{firstpage}
\pagerange{\pageref{firstpage}--\pageref{lastpage}}
\maketitle

\begin{abstract}
The importance of stellar rotation in setting the observed properties of young star clusters has become clearer over the past decade, with rotation being identified as the main cause of the observed extended main sequence turn-off (eMSTO) phenomenon and split main-sequences. Additionally, young star clusters are observed to host large fractions of rapidly rotating Be stars, many of which are seen nearly equator-on through decretion disks that cause self-extinction (the so called ``shell stars''). Recently, a new phenomenon has been reported in the $\sim1.5$~Gyr star cluster NGC 1783, where a fraction of the main sequence turn-off stars appears abnormally dim in the UV.  We investigate the origin of these ``UV-dim'' stars by comparing the UV colour-magnitude diagrams of NGC~1850 ($\sim100$~Myr), NGC~1783 ($\sim1.5$~ Gyr), NGC~1978 ($\sim2$~Gyr) and NGC~2121 ($\sim2.5$~Gyr), massive star clusters in the Large Magellanic Cloud. While the younger clusters show a non-negligible fraction of UV-dim stars, we find a significant drop of such stars in the two older clusters. This is remarkable as clusters older than $\sim$2 Gyr do not have an eMSTO, thus a large populations of rapidly rotating stars, because their main sequence turn-off stars are low enough in mass to slow down due to magnetic braking. 
We conclude that the UV-dim stars are likely rapidly rotating stars with decretion disks seen nearly equator-on (i.e., are shell stars) and discuss future observations that can confirm or refute our hypothesis.
\end{abstract}




\section{Introduction}
\label{sec:intro}
The star cluster environment offers a unique laboratory to observe peculiar stars and phenomena that are either not found or harder to study in the surrounding field. 
For example, star clusters allow us to investigate fascinating stellar exotica, such as white dwarfs, nova remnants and black holes (e.g. \citealt{giesers18,goettgens19}).
It is also very well known that massive star clusters host star-to-star chemical abundance variations, down to an age of $\sim1.5$ Gyr (see e.g. \citealt{bastianlardo18} and \citealt{cadelano22}), which are not commonly found in field stars (e.g. \citealt{martell11}). Star clusters younger than $\sim1.5$ Gyr are instead not found to show such multiple stellar populations, at least not in their evolved stars (e.g. \citealt{cabreraziri16b,lardo17}).

On the other hand, young and intermediate age star clusters (a few Myr up to $\sim$2 Gyr old) have been found to host other peculiar features in their colour-magnitude diagrams (CMDs) that manifest in the form of extended main sequence turn-offs (eMSTOs) or split/dual main sequences (MSs, e.g. \citealt{bertelli03,mackey07,milone09,girardi13,milone18,milone22}). Interestingly, such features are observed in massive clusters ($\gtrsim10^4$\msun) in the Magellanic Clouds (e.g. \citealt{goudfrooij14,milone15,bastian16}) as well as in the less massive ($\lesssim10^4$\msun) open clusters in the Milky Way ($\lesssim10^3$\msun , \citealt{bastian18,cordoni18}), suggesting their presence is ubiquitous and does not depend on the environment. These cannot be explained by spreads in age (e.g. \citealt{cabrera-ziri16,cordoni22}), and they appear to be fundamentally different from the phenomenon of multiple populations mentioned above. Indeed, the properties of the eMSTO are strongly correlated with the age of the clusters, thus suggesting a stellar evolutionary effect is the cause \citep{niederhofer15}.

Recent works have shown that the eMSTO and split MSs in star clusters are due to a single age stellar population with a range of rotation rates (e.g. \citealt{bastiandemink09,brandthuang15,dantona15,dupree17,kamann18,marino18,kamann20}). However, the origin of such different rotational rates is still under debate (e.g. \citealt{bastian20, wang22}).
Understanding how this phenomenon works is crucial not only for star cluster formation studies, but also to set fundamental constraints on stellar evolution models that aim at including stellar rotation effects. 


Interestingly, \cite{milone22}, following \cite{milonemarino22}, have recently unveiled the existence of an additional peculiarity on the MSTO of NGC 1783 ($\sim$1.5 Gyr old star cluster in the Large Magellanic Cloud, LMC). They found that some turn-off (TO) stars ($\sim$7\%) are shifted to the red in the F275W-F438W vs F814W HST CMD. They named these peculiar stars ``UV-dim'' stars. The authors explored the possibility that stellar rotation might be the underlying cause of the extreme red colours of these stars. They made a comparison of the data with simulations, by using isochrones that also include stellar rotation. They conclude that it is quite unlikely that the
observed spread of the UV-dim stars can be entirely due to stars rotating close to break$-$up velocities at the age of NGC~1783. They also mentioned that circumstellar disks might affect the UV radiation of the UV-dim stars.
Hence, the physical nature of these stars is currently unknown.

\begin{figure*}
\centering
\includegraphics[scale=0.47]{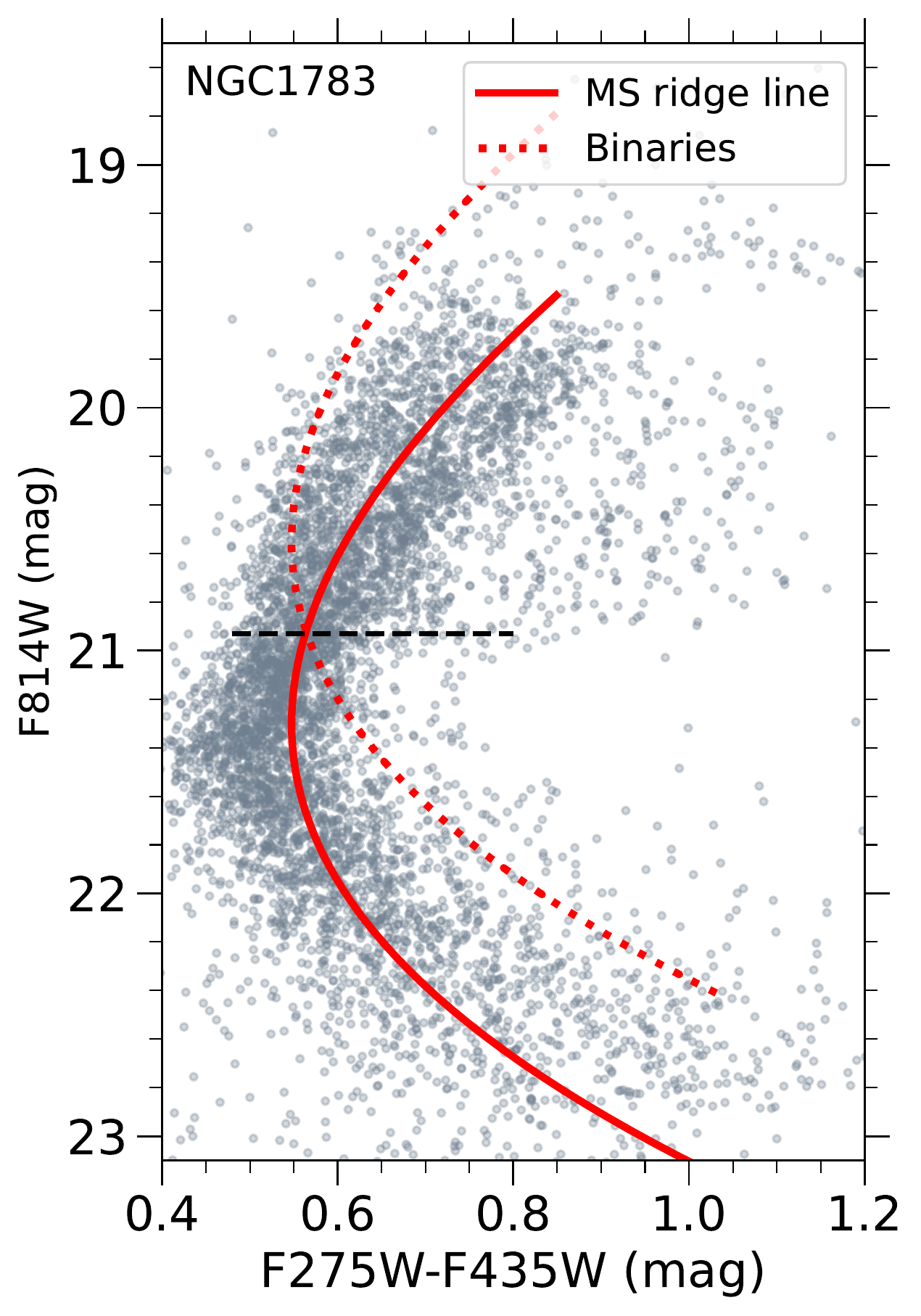}
\includegraphics[scale=0.47]{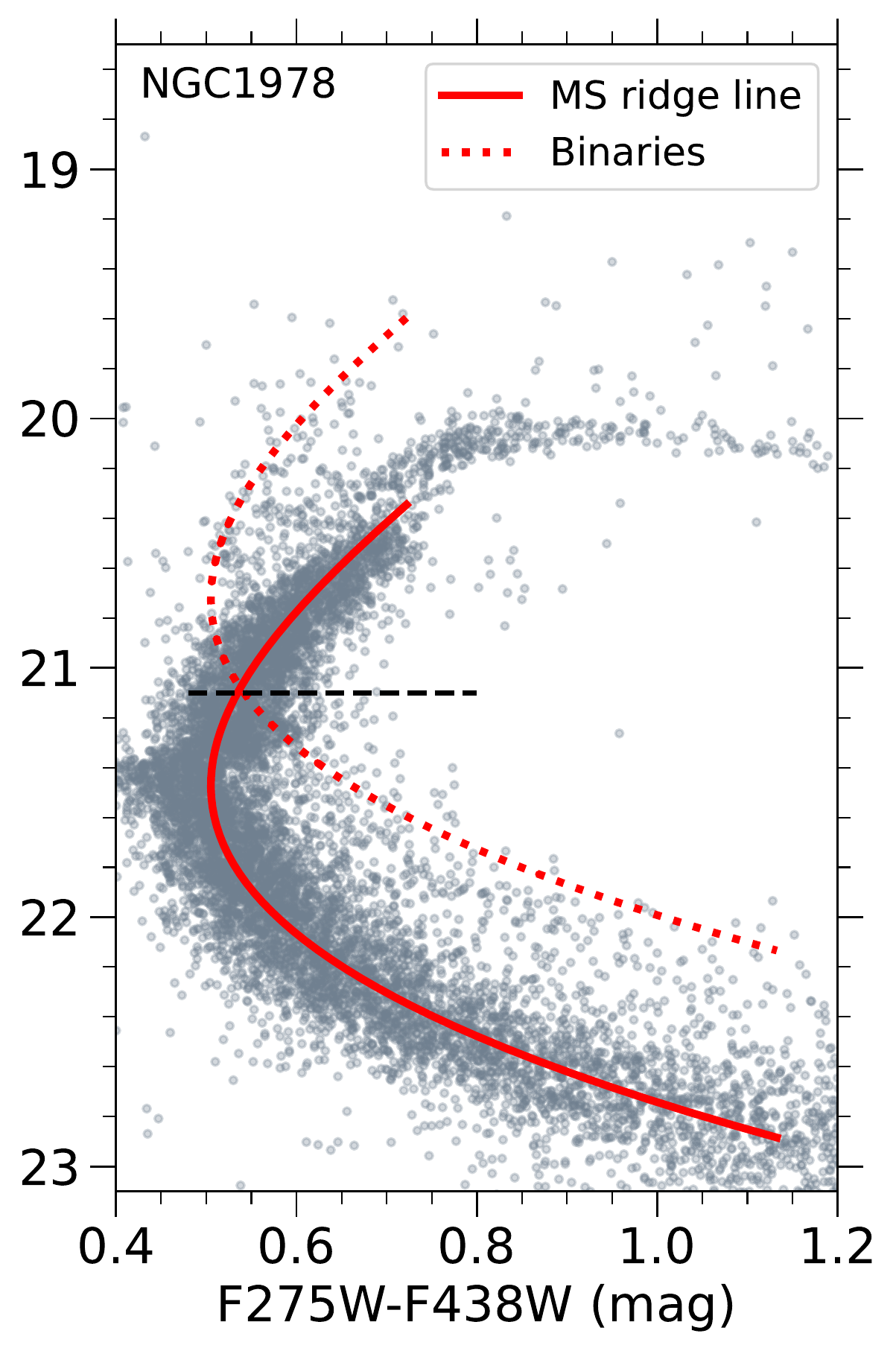}
\includegraphics[scale=0.47]{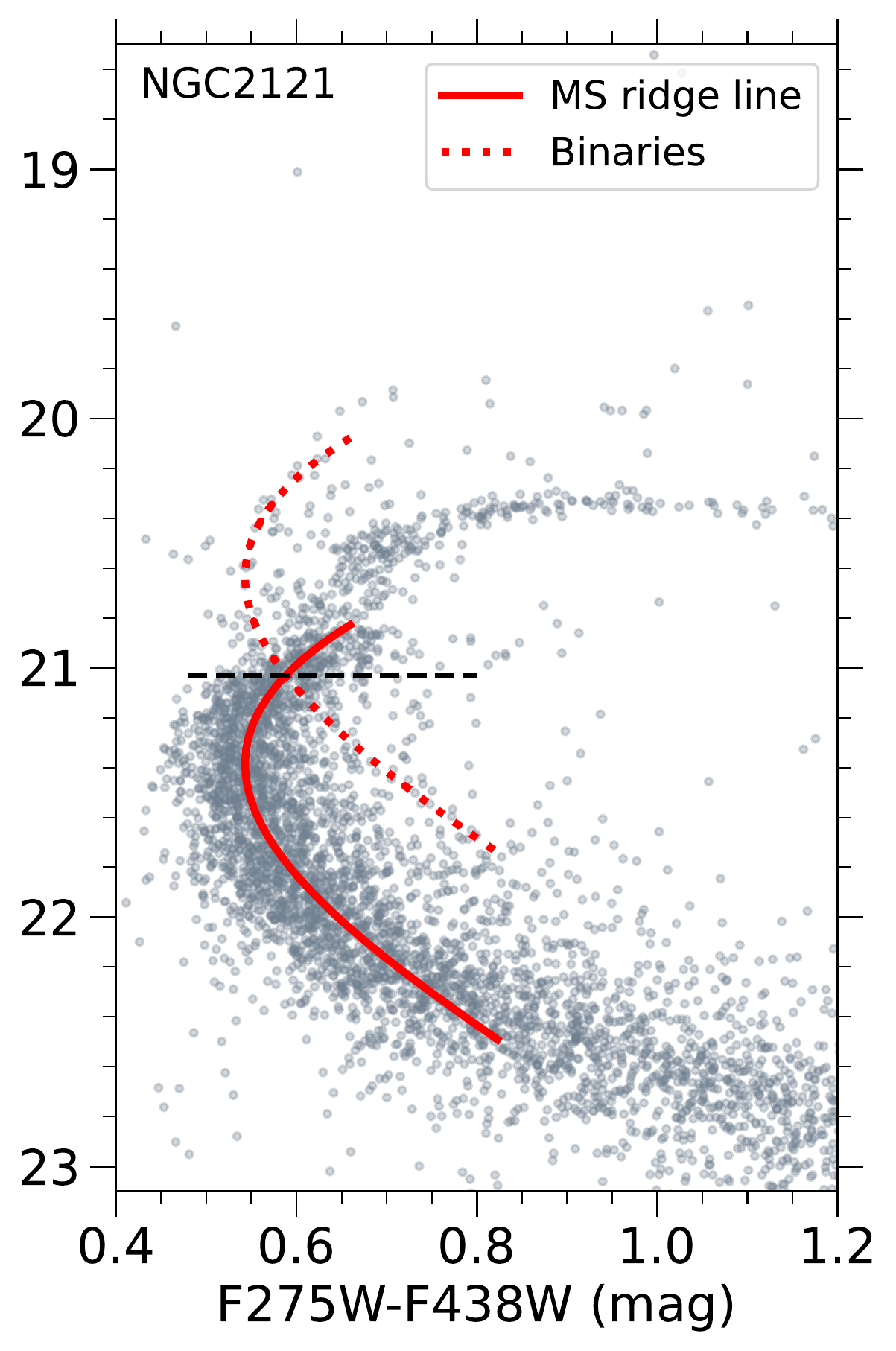}
\caption{CMDs in F275W-F438/5W vs F814W HST filters for the intermediate age clusters in the sample, namely NGC~1783, NGC~1978 and NGC~2121 from left to right, respectively. The red solid line indicates the ridge line calculated over the MS range. The dotted line represents the loci where equal-mass,  pre-interaction binaries are expected to lie. The black dashed horizontal line indicates the faintest edge for the selection of turn-off and UV-dim stars, adopted to limit the contribution of binaries to the selection. See text for more details.}
\label{fig:binaries}
\end{figure*}

A possibly similar phenomenon has been observed in the $\sim$100 Myr massive star cluster NGC~1850 in the LMC. 
Very recently, \cite{kamann22} studied the MS of NGC~1850 with HST photometry and MUSE spectroscopy. 
They measured the projected rotational velocities in the MS of the cluster, confirming that stars that are on the red edge of the MSTO are fast rotators, while stars on the blue edge are slow rotators (see their Fig. 3). Additionally, they found that $\sim$50\% of stars in the MSTO region are Be stars, B stars that are rapidly rotating and have emission lines due to their decretion disks (e.g. \citealt{feast72,grebel92}). Such Be stars have been detected on the MSTO of a number of clusters in the Magellanic Clouds at different ages (from $\sim$40 up to a few 100s Myr, e.g. \citealt{bastian17,milone18,bodensteiner20}), thus representing even more evidence of the impact of stellar rotation in shaping the CMDs of such clusters. A fraction of the Be stars in NGC~1850 ($\sim$23\%, \citealt{kamann22}) shows shell-like features, such as a double peak H$\alpha$ line, deeper Paschen and narrow FeII and SiII lines.  These stars are named shell stars, i.e. Be stars observed nearly equator-on and thus extincted by their own disk (e.g. \citealt{rivinius06}). Indeed, in the HST F336W-F438W vs F438W CMD of NGC~1850 (see Fig. 10 of \citealt{kamann22}), the shell stars are found to be shifted to the red with respect to the bulk of MSTO stars, similarly to what it is observed for the UV-dim stars in NGC~1783.

In this work, we aim at shedding light on the origin of UV-dim stars by investigating the UV CMDs of clusters of different ages.
According to the stellar rotation paradigm, it is expected that the eMSTO disappears after a cluster reaches an age of $\sim$2 Gyr (e.g. \citealt{georgy19}), as stars on the turn-off at this age are 
enough low in mass (masses $\lesssim 1.5$~\msun) to have convective envelopes. Consequently, they have surface magnetic fields which can brake the star. As the stars are braked, no rapid stellar rotators are expected to be present after this age.
Observations nicely show that clusters aged $\sim$2 Gyr do not show an eMSTO \citep{martocchia18b,yong22}. If the UV-dim phenomenon is linked to shell stars (and thus stellar rotation), we should expect to see UV-dim stars predominantly in clusters that contain rapid rotators, i.e., those with ages below 2 Gyr.

In this paper, we report the measurement of the fractions of UV-dim stars in the MSTO of four star clusters in the LMC (namely NGC~1850, NGC~1783, NGC~1978 and NGC~2121), for which UV HST photometry is available.
These objects are respectively aged $\sim100$ Myr and $\sim$1.5, $\sim$2 and $\sim$2.5 Gyr (e.g. \citealt{niederhofer16, martocchia18b,martocchia19}), thus straddle the age limit where rapidly rotating stars are, or are not, expected. 

This paper is organised as follows. Section \ref{sec:data} describes the data used for the analysis, which is outlined in Section \ref{sec:analysis}. In Section \ref{sec:res} we report on the results, while we discuss and conclude in Section \ref{sec:disc}.

\section{Data and observations}
\label{sec:data}
The data used in this paper consist of HST photometric catalogues of the LMC star clusters NGC~1850, NGC~1783, NGC~1978 and NGC~2121, previously published by \cite{saracino20a,saracino20b} and \cite{kamann21}. 
The observations for NGC~1783 and NGC~2121 are from several GO HST programmes that are reported in Table 1 of \cite{saracino20a}. They include the filters F275W, F336W, F343N, F435W(F438W) and F814W in both UVIS/WFC3 and ACS. Data for NGC~1978 are from GO-14069 and GO-15630 (PI: N. Bastian) and consist of F275W, F336W, F343N, F438W and F814W filters with the UVIS/WFC3 (see \citealt{saracino20b}). The photometric reduction of the intermediate age clusters has been performed using \texttt{DAOPHOT IV} \citep{stetson87}, and the cross-correlation software \texttt{CataPack} \citep{montegriffo95}. The procedure is extensively described in the above mentioned papers \citep{saracino20a,saracino20b} as well as in \cite{martocchia18a}. Data for NGC~1850 also consist of different HST GO programmes (GO-14069, PI: N. Bastian and GO-14174 PI: P. Goudfrooij) with images in filters F275W, F336W, F343N, F438W,
F656N, and F814W, taken with the WFC3. The photometric analysis has been performed with \texttt{DOLPHOT} \citep{dolphin2000} and all the steps are reported in \cite{gossage19} as well as in \cite{kamann21}.
We refer the interested readers to all these references for more details on the photometric analysis.

Cluster stars are selected within 40 arcsec from the cluster centre for NGC~1978 and NGC~2121 (45 arcsec for NGC~1783), where field interlopers are removed through a statistical approach. Such a field stars decontamination method was also quite extensively described in \cite{saracino20a} and \cite{cabrera20}. For NGC~1850, a field star decontamination was not performed because the cluster's turn-off stars are much brighter than the LMC field stars and thus contamination is negligible, at the magnitude level we are interested in to look for UV-dim stars. Finally, the effect of differential reddening (DR) on the final catalogues has been checked. DR was already estimated for these clusters in \cite{saracino20a,saracino20b}, according to the method reported in \cite{milone12}. The obtained $\delta E(B-V)$ values are quite low for all the clusters (on average around zero and with a maximum variation comparable to the photometric errors), hence we did not correct the catalogues used in this work for DR effects.

\section{Analysis}
\label{sec:analysis}

\subsection{NGC~1783, NGC~1978, NGC~2121}
\label{subsec:analysis-int}

In this Section, we report on the analysis of the intermediate age clusters of the sample. For NGC~1850, we use a slightly different method which we will describe in Section \ref{subsec:analysis-1850}.

Figure \ref{fig:binaries} shows the HST CMDs of NGC~1783 ($\sim1.5$~Gyr), NGC~1978 ($\sim2$~Gyr) and NGC~2121 ($\sim2.5$~Gyr), in the F275W-F438W(or F435W) vs. F814W space. We chose the F275W UV filter for our analysis, as it is the available filter with the shortest wavelength, hence the most effective at separating UV-dim stars from 'normal' TO stars (see \citealt{milone22}).
From a first visual inspection of Fig. \ref{fig:binaries}, many stars around the TO of NGC~1783 (with magnitudes $19.7\lesssim$F814W$\lesssim 21.0$ mag) show very red colours, and therefore they appear shifted to the right of the MSTO. Throughout the paper, we will refer to these stars as UV-dim stars, as defined in \cite{milone22}. Furthermore, these stars seem to disappear in the two older clusters, NGC~1978 and NGC~2121 (age $\geq$2 Gyr).
However, in order to demonstrate this trend, we need to select our bona-fide UV-dim stars.

First, we defined a ridge line over the MS of each cluster in the F275W-F438/5W vs. F814W CMD, by using a smoothing spline fit with third-degree polynomials. This is displayed as a red solid curve in all panels of Fig. \ref{fig:binaries}. From this ridge line, we constructed a second curve, which is shown as a dotted red curve in Fig. \ref{fig:binaries}. Such a curve indicates where the equal-mass MS binaries are expected to lie in the CMD, before an interaction occurs. 
Indeed, if there are two equal-mass MS stars in a pre-interaction binary system, they have the same colour but twice the luminosity compared to a single MS star of that mass.
By converting this factor of 2 in luminosity into magnitude, we obtained the equation for the equal mass, pre-interaction, binaries (eqm, binaries): $m_{\rm eqm, binaries\, r.l.}=m_{\rm r.l.}-0.75$ mag, where r.l. stands for ridge line. This equation corresponds to the dotted red curves plotted in Fig. \ref{fig:binaries}. We defined the faintest edge of our selection of UV-dim stars as the intersection of the MS ridge line and the equal mass binaries ridge line, for each intermediate age cluster. This threshold is indicated as a black dashed horizontal line in Fig. \ref{fig:binaries}. In this way, we minimised the presence of pre-interaction binaries in our UV-dim stars samples. Additionally, we note that post-interaction MS binaries, due to mergers, are not expected to contaminate our samples of UV-dim stars, as their colours are predicted to be bluer than the MSTO in UV CMDs (see e.g. Figure 3 of \citealt{wang22}).

\begin{figure*}
\centering
\includegraphics[scale=0.44]{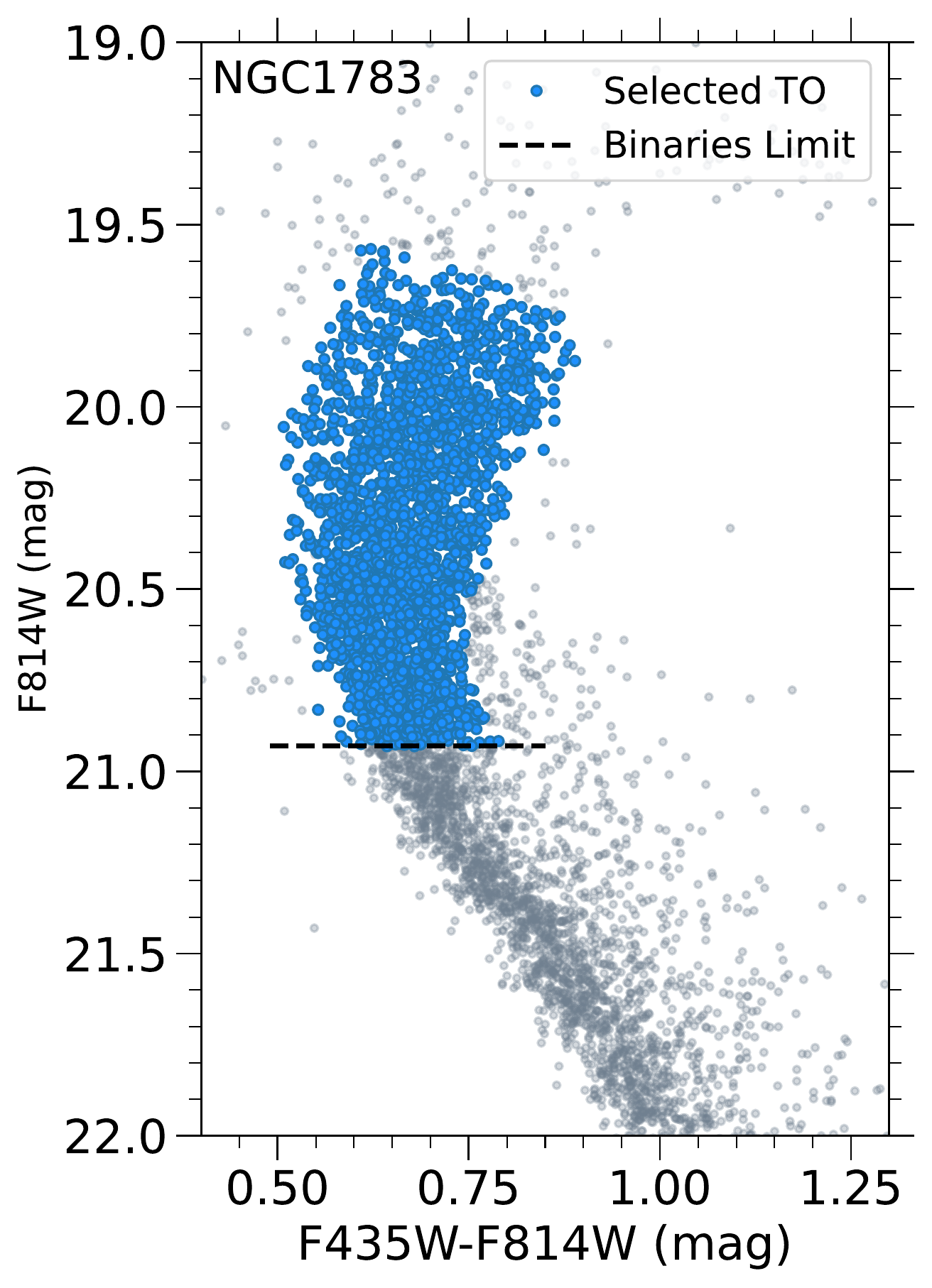}
\includegraphics[scale=0.44]{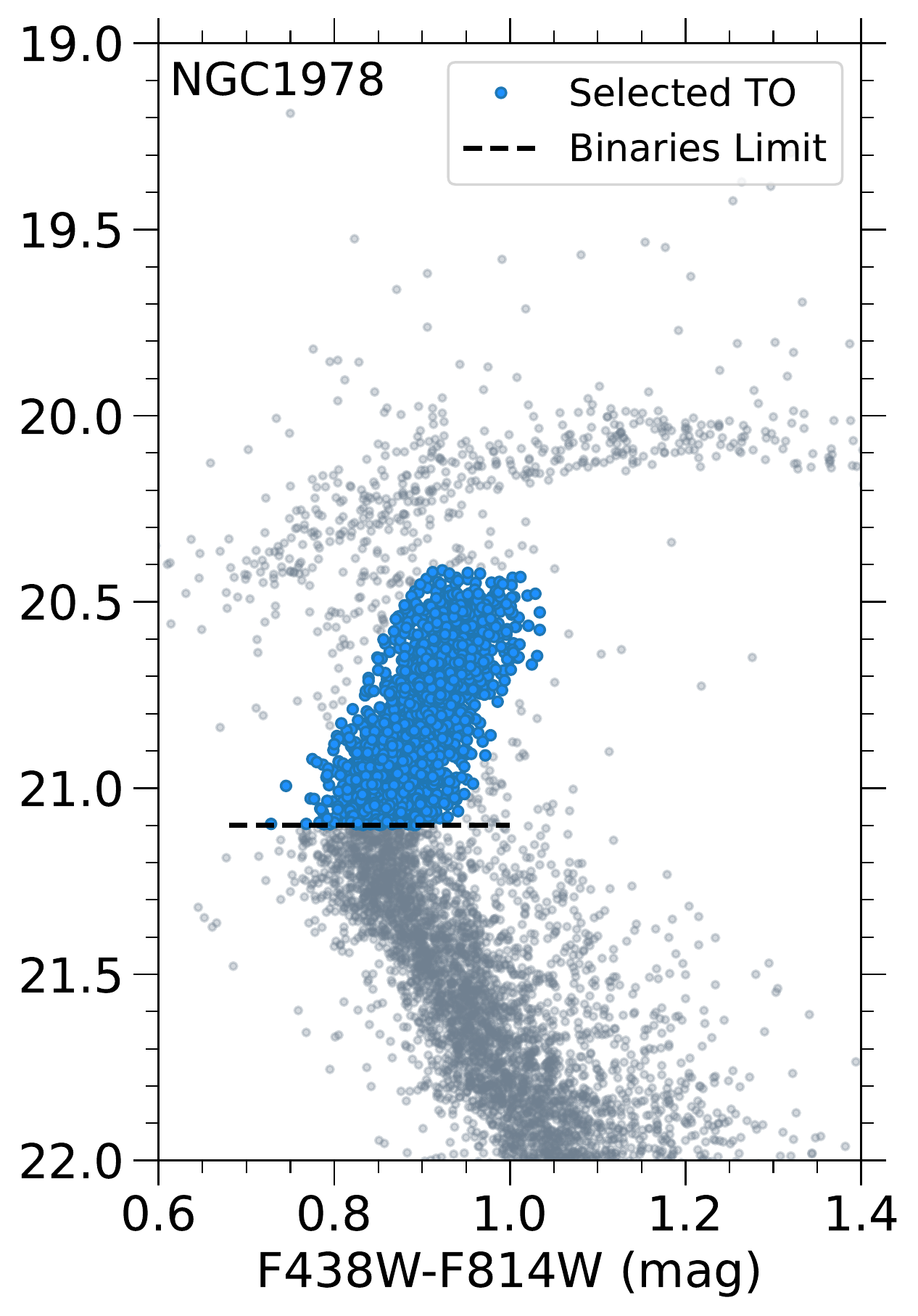}
\includegraphics[scale=0.44]{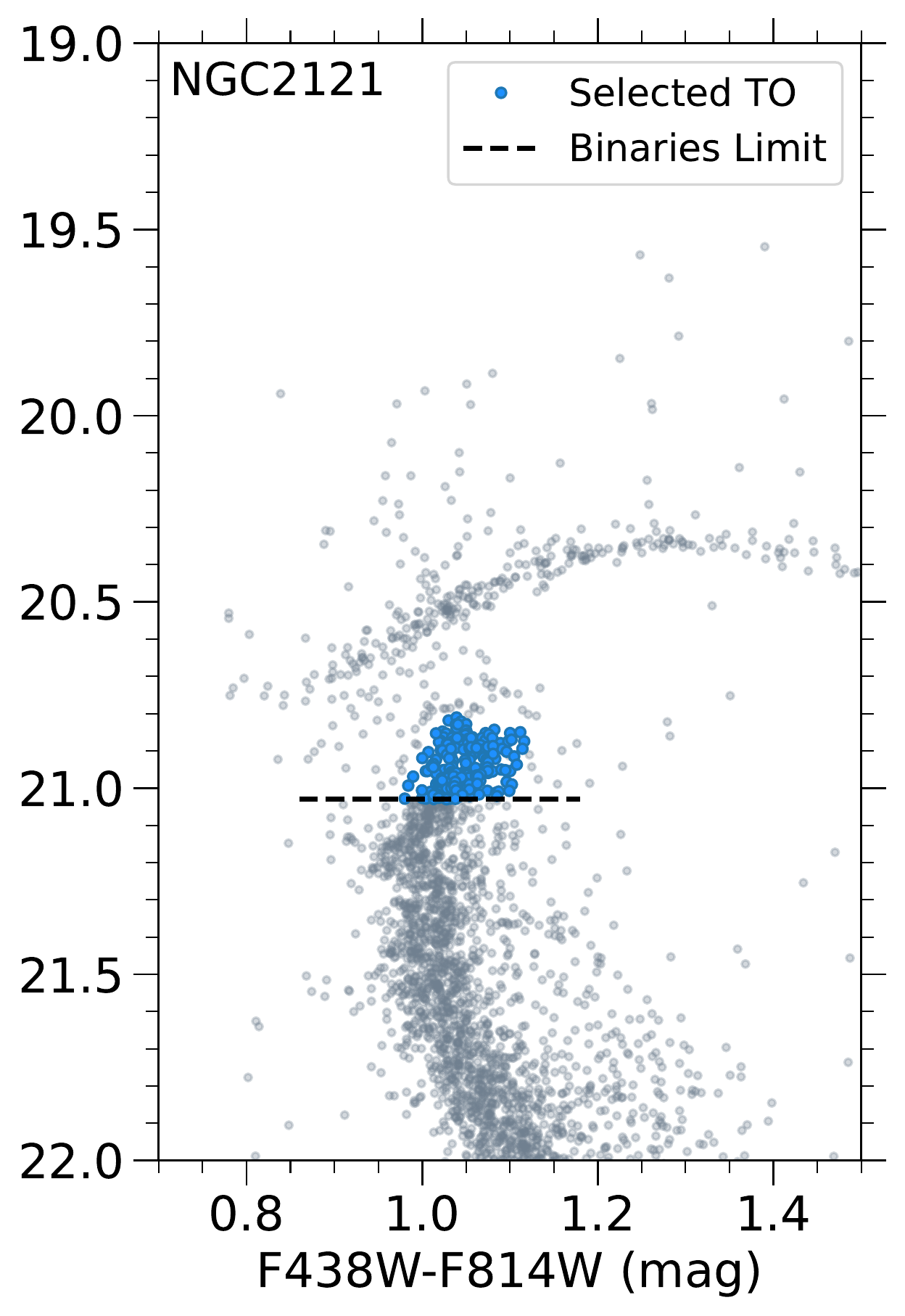}
\caption{CMDs in F438/5W-F814W vs F814W HST filters for the intermediate age clusters in our sample. The light blue circles represent the selected TO stars in optical colours. The black dashed horizontal line indicates the faintest edge for the selection, previously defined from Fig. \ref{fig:binaries}.}
\label{fig:TO}
\end{figure*}

\begin{figure*}
\centering
\includegraphics[scale=0.5]{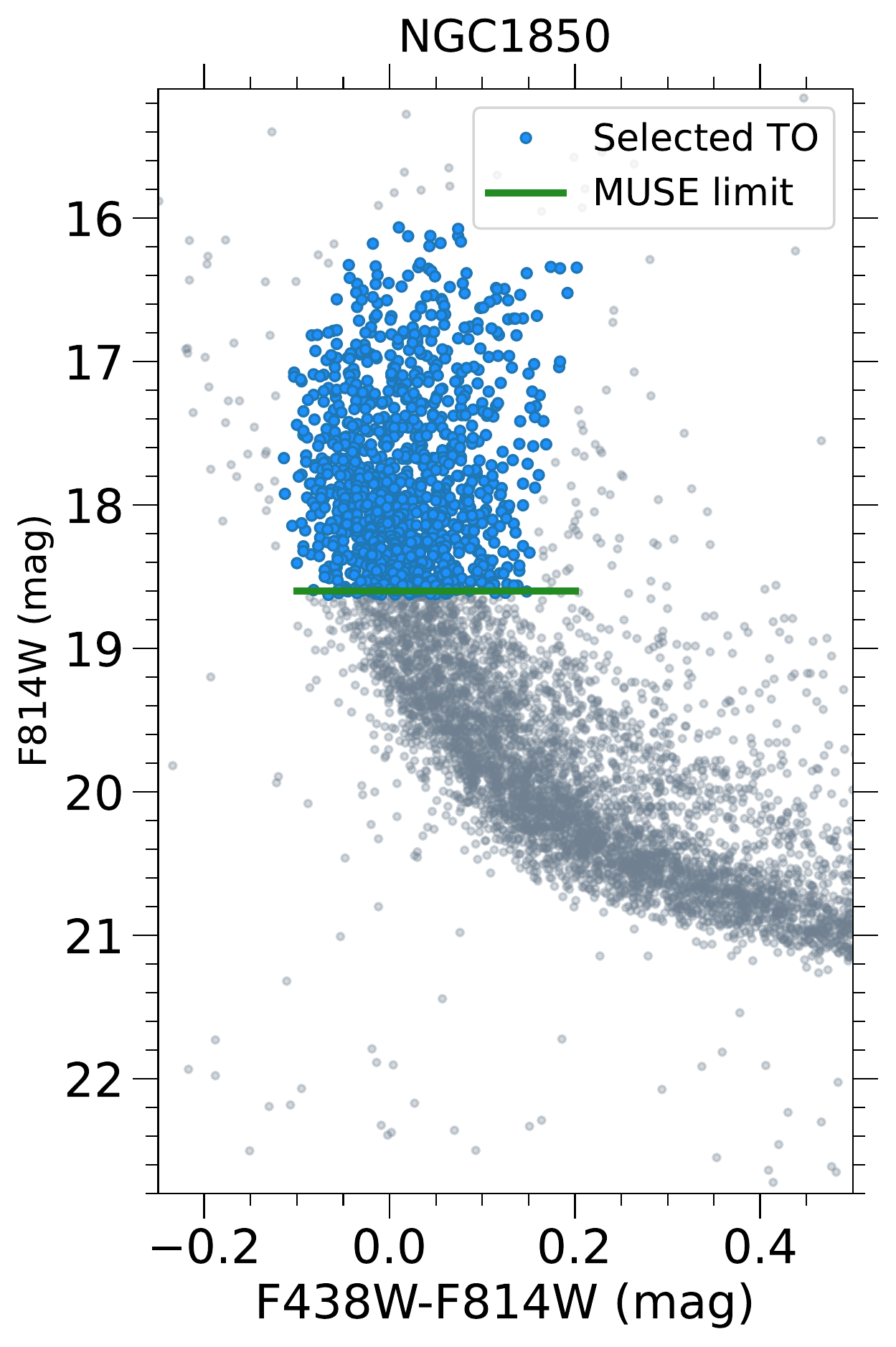}
\includegraphics[scale=0.5]{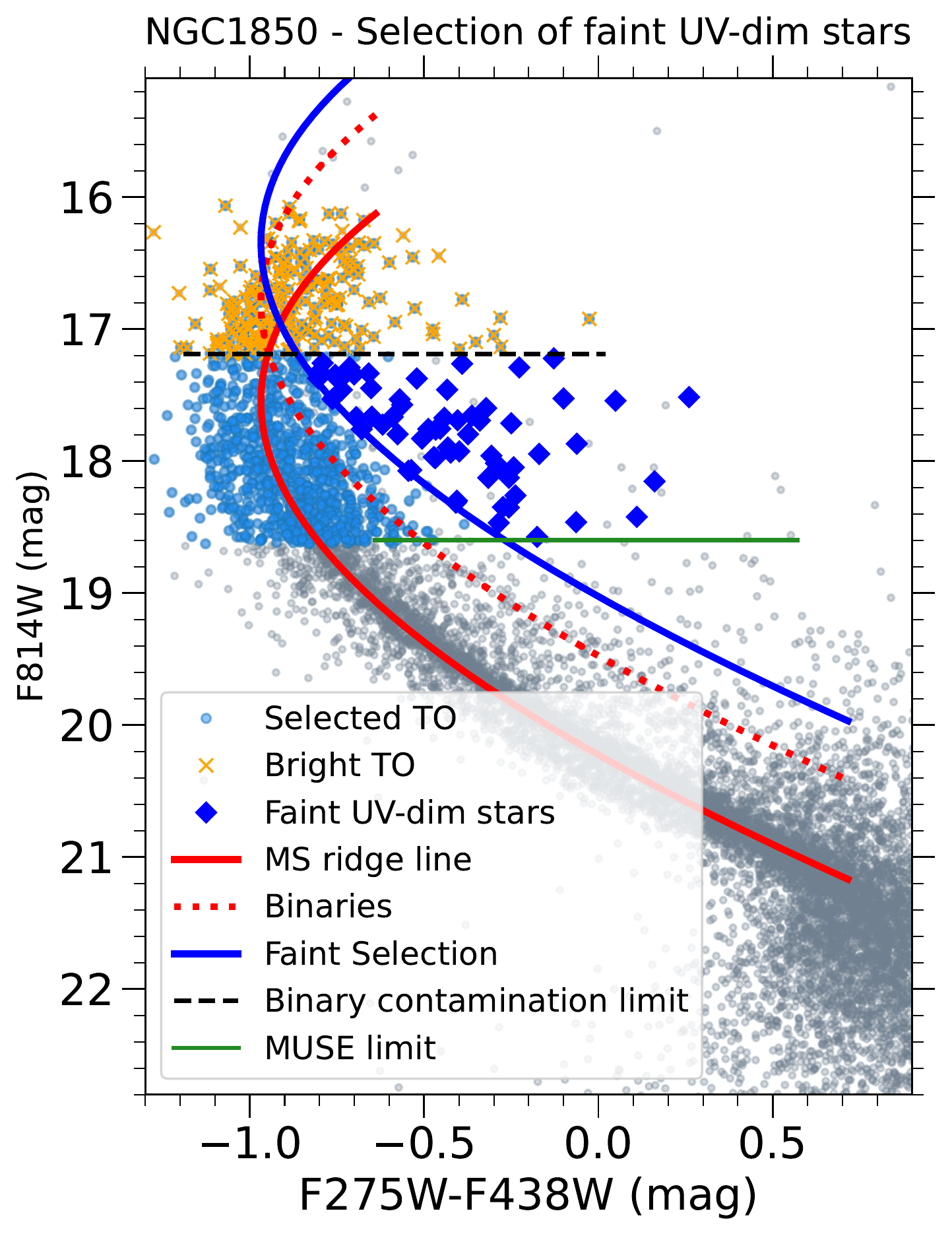}
\caption{\textit{Right panel}: CMD in F438W-F814W vs F814W HST filters of NGC~1850. The light blue circles represent the selected TO stars in this optical colour, while the green horizontal line indicates the magnitude limit at which shell and Be stars are identified spectroscopically with MUSE \protect\citep{kamann22}. \textit{Left panel}: CMD in F275W-F438W vs F814W HST filters of NGC 1850. The red solid line indicates the ridge line calculated over the MS range. The dotted line represents the loci where equal-mass pre-interaction binaries are expected to lie. The black dashed horizontal line indicates the faintest limit where binaries start to significantly contribute to the red part of the MS. Light blue circles represent the total selected TO stars (from the left panel), while the orange crosses indicated the selected bright TO, from which bright UV-dim stars are selected. The blue solid curve represent a shifted MS ridge line used for the selection of faint UV-dim stars (indicated as blue diamonds). See text for more details.}
\label{fig:analysis-1850}
\end{figure*}

As the optical colours are not sensitive to the UV emission of our UV-dim stars candidates, we used them to select a clean sample of MSTO stars. Figure \ref{fig:TO} shows the HST optical CMDs of NGC~1783, NGC~1978 and NGC~2121, respectively from left to right, in the F438W(or F435W)-F814W vs. F814W space. 

To select our bona-fide TO stars, we used three steps. The first step (1) is reported above, where we defined a faint edge for the selection of TO stars from Fig. \ref{fig:binaries} (black dashed horizontal line). In a second step (2), we selected TO stars by hand in the turnoff region of the intermediate age clusters in the optical CMDs (see light blue circles in Fig. \ref{fig:TO}). 
Finally (step 3), we plotted these selected stars again in the UV CMDs (see next Section \ref{sec:res}). In this way, we ensure that we considered only those stars that are located in the TO region and are not scattered around other parts of the CMD.
In the next Section \ref{sec:res}, we will then use these samples in order to select our UV-dim stars.

\subsection{NGC~1850}
\label{subsec:analysis-1850}

In this Section, we report on the analysis of NGC~1850, the youngest cluster in the sample ($\sim$ 100 Myr old). We used a different method for this cluster for two main reasons: (i) NGC~1850 is much younger than the other clusters in the sample, and (ii) spectroscopic information for its MSTO stars is available (\citealt{kamann22}). Hence, we already know that NGC~1850 hosts many rapidly rotating Be stars, a fraction of which are shell stars that are red in F336W-F438W vs F438W colours \citep{kamann22}. Therefore, we will not limit our analysis to the brightest part of the TO which is poorly contaminated by binaries (as done for the other clusters, see Sections \ref{subsec:analysis-int}), but we will also select photometric UV-dim stars down to the MS of the cluster. As MUSE spectroscopy is available for NGC~1850, we will make a comparison between UV-dim stars and shell/Be stars by using MUSE spectra\footnote{Available at the website: \url{https://cdsarc.cds.unistra.fr/viz-bin/cat/J/MNRAS/518/1505}}, in Section \ref{sec:res}.

The left panel of Fig. \ref{fig:analysis-1850} shows the F438W-F814W vs F814W CMD of NGC~1850, where light blue circles represent the selected TO stars. The faintest limit for the selection is indicated as a green horizontal line in the left panel of Fig. \ref{fig:analysis-1850} and it represents the faint limit at which shell and Be stars are identified with MUSE spectroscopy.

The right panel of Figure \ref{fig:analysis-1850} shows the F275W-F438W vs F814W CMD of NGC~1850, where a similar analysis as in Fig. \ref{fig:binaries} has been performed. Indeed, the red solid line indicates a ridge line defined over the MS of the cluster, while the red dotted curve represents the equal-mass, pre-interaction, binary line for NGC~1850 in this colour combination (see the previous Section \ref{subsec:analysis-int}). The black dashed horizontal line represents the magnitude where these two curves intersect and it represents the faintest limit where pre-interaction binaries do not contribute significantly to the red edge of the TO. The light blue circles represent the total selected MSTO stars from the left panel of Fig. \ref{fig:analysis-1850}. We divided the total selected TO stars into a ``bright'' and ``faint'' sample of stars. Bright TO stars have magnitudes brighter than the binary contamination limit (black dashed line) and are indicated as orange crosses in Fig. \ref{fig:analysis-1850}. 
Faint TO stars are defined as the stars fainter than the binary contamination limit and brighter than the MUSE limit (green solid line in Fig. \ref{fig:analysis-1850}). We will select UV-dim stars in NGC~1850 from both these samples, in a different way.

To limit the (pre-interaction) binary contamination in the faint TO sample, we defined a ridge line that is defined as the MS ridge line $-1.2$ mag, indicated as a blue solid curve in Fig. \ref{fig:analysis-1850}. 
Thus, we selected the UV-dim stars in the faint TO sample (named as faint UV-dim stars) having F275W-F438W colours redder than the blue curve in Fig. \ref{fig:analysis-1850}. Faint UV-dim stars are indicated as blue diamonds in the right panel of Fig. \ref{fig:analysis-1850}. The choice of the blue curve for the selection of faint UV-dim stars is made by eye, defined to be $\sim$0.5 mag more luminous than the pre-interaction binary contamination line (red dotted curve). By looking at the distribution of stars with colours redder than the red dotted curve, the blue curve represents a rough indication of the colours where there is a drop in stellar density, thus we can expect a lower contamination from pre-interaction binaries here.\footnote{We also calculated the fraction of UV-dim stars when using a different threshold for the faint selection. We used a new curve that is defined as the MS ridge line $-1.5$ mag to select faint UV-dim stars. In this case, the total number of UV-dim stars in NGC~1850 slightly decreases, however the conclusions of the paper stay unchanged.}

For the bright TO sample of NGC~1850, we select the bright UV-dim stars in the same way as for the other clusters. We will calculate the  final fraction of UV-dim stars in NGC~1850 by combining the faint and bright selected samples of UV-dim stars in the next Section \ref{sec:res}, where we will also report on the results for all the clusters.

\begin{figure*}
\centering
\includegraphics[scale=0.5]{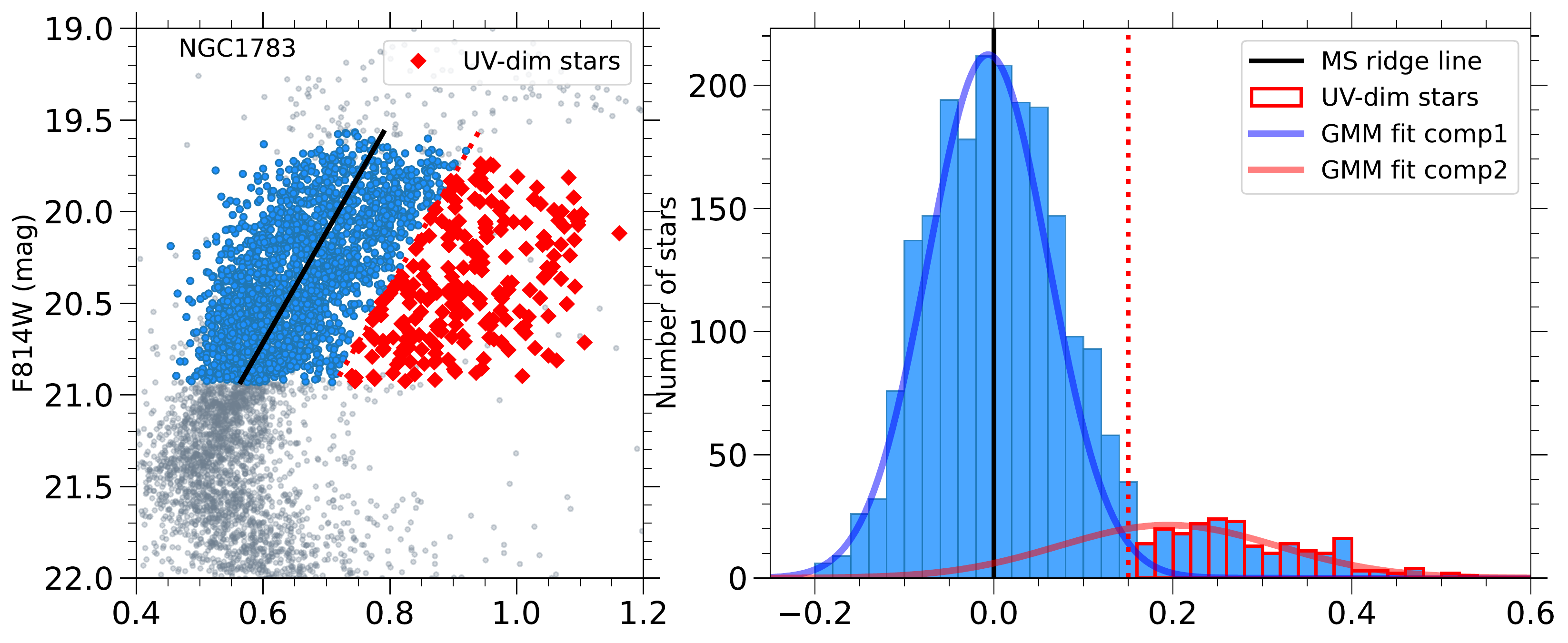}
\includegraphics[scale=0.5]{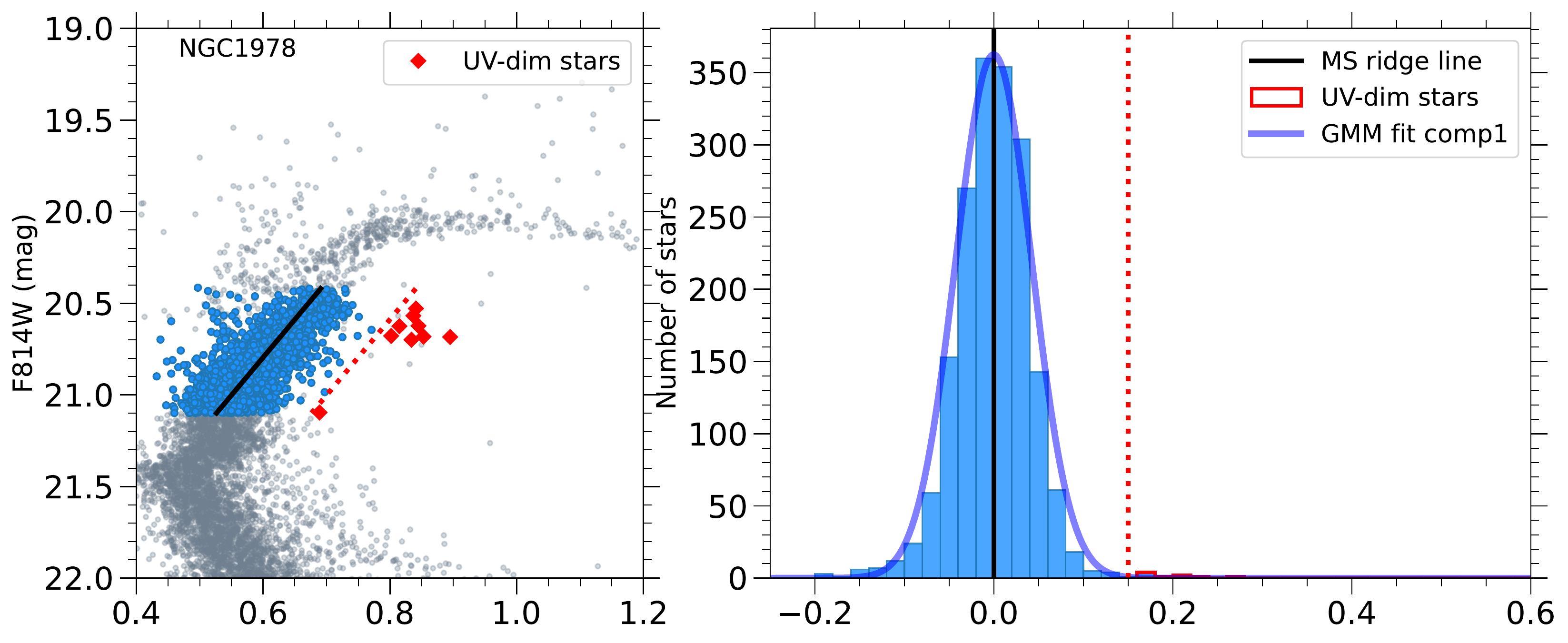}
\includegraphics[scale=0.5]{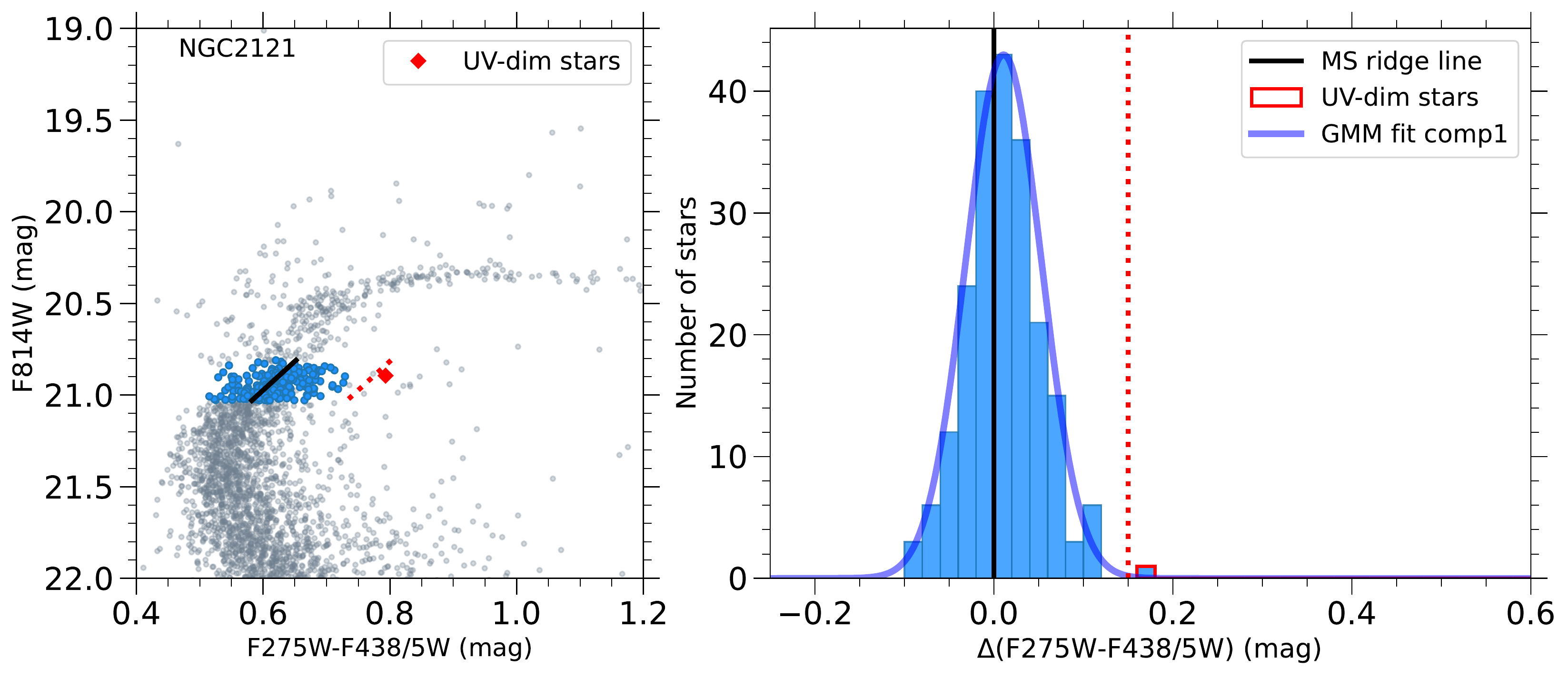}
\caption{\textit{Left panels}: F275W-F438/5W vs F814W HST CMDs of the three intermediate age clusters of the sample. Blue circles represent the previously selected TO stars (Fig. \ref{fig:TO}), while the black solid line is the MSTO ridge line. The red diamonds indicate UV-dim stars selected from the right panels. \textit{Right Panels:} Histograms of the distribution of $\Delta$(F275W-F438/5W) verticalised colours of TO stars. The red dotted vertical line indicates the selection threshold in the distribution (i.e., $\Delta$(F275W-F438/5W)$=$0.15 mag, see text) for the UV-dim stars (in red), also reported unverticalised in the left panels. Blue (red) gaussian curves indicate the respective GMM fit components.}
\label{fig:uv-dim}
\end{figure*}

\begin{figure*}
\centering
\includegraphics[scale=0.3]{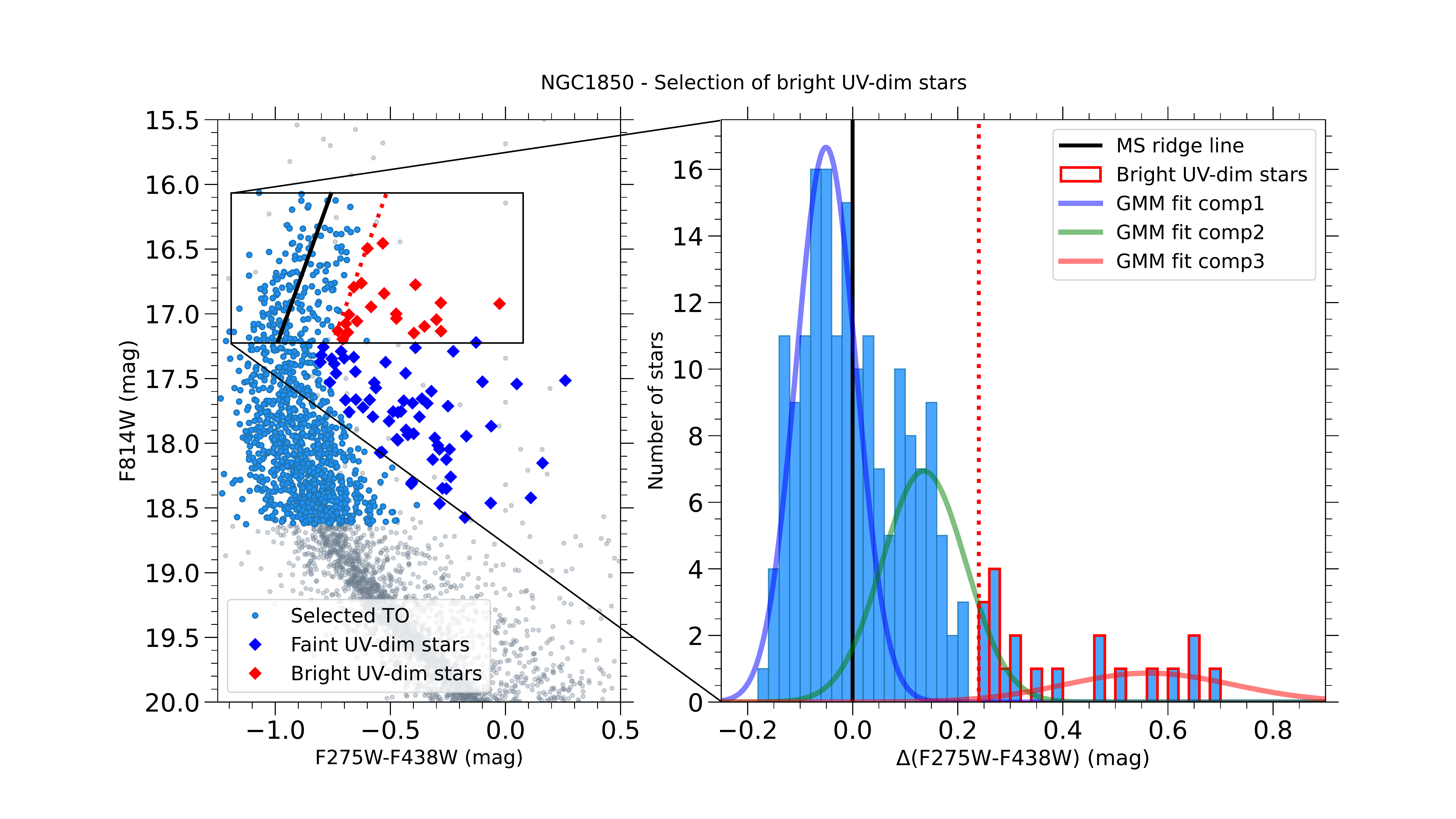}
\caption{As in Fig. \ref{fig:uv-dim} but for NGC~1850. The selection threshold for bright UV-dim stars is $\Delta$(F275W-F438/5W)$=$0.24 mag, see text for more details. Blue diamonds indicate the selected faint UV-dim stars from Fig. \ref{fig:analysis-1850}.}
\label{fig:uv-dim-1850-bright}
\end{figure*}

\section{Results}
\label{sec:res}

In this Section, we estimate the fractions of UV-dim stars in all the clusters of the sample. 

In the left panels of Figures \ref{fig:uv-dim} and \ref{fig:uv-dim-1850-bright}, we plotted the selected TO stars (from Figs. \ref{fig:TO} and \ref{fig:analysis-1850}, light blue circles) in the HST UV CMDs of NGC~1783, NGC~1978 and NGC~2121, and NGC~1850, respectively, in F275W-F438/5W vs. F814W space. Next, we defined a new MS ridge line on the selected TO stars\footnote{For NGC~1850, we consider only the bright TO stars here, indicated as a black box in Fig. \ref{fig:uv-dim-1850-bright}, see Section \ref{subsec:analysis-1850}.} (solid black line in Figs. \ref{fig:uv-dim} and \ref{fig:uv-dim-1850-bright}), by using a smoothing spline fit of first-degree polynomials\footnote{The new first-degree spline used here visually fits the data better with respect to a third degree spline, given the shape of the selected stars in the CMDs. However, we checked that the choice of the spline does not affect our results and thus the estimated fractions of UV-dim stars.}. Then, we calculated the distance of each TO star from the ridge line to obtain the $\Delta$(F275W-F438/5W) verticalised colours. Finally, the right panels of Figures \ref{fig:uv-dim} and \ref{fig:uv-dim-1850-bright} show the histograms of the distribution of the verticalised $\Delta$(F275W-F438/5W) colours. Just from a visual inspection, we noticed that the distributions of NGC~1783 and NGC~1850 (right top panel of Fig. \ref{fig:uv-dim}, right panel of Fig. \ref{fig:uv-dim-1850-bright}) are multimodal, compared to the unimodal distributions of NGC~1978 and NGC~2121. More quantitatively, we fit the $\Delta$ colour distributions of each cluster with Gaussian Mixture Models (GMMs) to identify the presence of multiple gaussian components. We used the \texttt{SCIKIT-LEARN} \citep{scikit-learn} python package called \texttt{MIXTURE}\footnote{\url{http://scikit-learn.org/stable/modules/mixture.html}}, which consists of an expectation-maximization algorithm for fitting mixtures of Gaussian models. As expected, for NGC~1783 the GMM fit found two Gaussian components, represented by the blue and red gaussian curves in the top right panel of Fig. \ref{fig:uv-dim}, respectively. For NGC~1978 and NGC~2121, the GMM fit yielded only one Gaussian component (represented by the blue gaussian curve in the middle and bottom panels of Fig. \ref{fig:uv-dim}). Finally, for NGC~1850, the GMM fit to the distribution of $\Delta$(F275W-F814W) found three components, with a second clearly defined peak and a third distribution that is skewed to higher values of $\Delta$(F275W-F438W). It is highly likely that the two peaks in the distribution are caused by the different rotation rates of blue and red MS stars (see also Fig. 3 from \citealt{kamann22}).

Next, we defined a threshold in $\Delta$(F275W-F438/5W) colour equal to 0.15 mag from the NGC~1783 distribution. This value has been chosen roughly to be the point where the two gaussian functions cross each other. This is shown as a red dotted vertical line in the right panels of Fig. \ref{fig:uv-dim} and it is the same for NGC~1978 and NGC~2121. Hence, UV-dim stars are selected to have $\Delta$(F275W-F438/5W)$>$0.15 mag and are displayed as red diamonds in the CMDs on the left hand panels of Fig. \ref{fig:uv-dim}. 

As NGC~1850 is much younger than the other clusters, its CMD shape is very different. Hence, we defined a different threshold for the selection of the bright UV-dim stars, with respect to the intermediate age clusters. This is also because both the second and the third gaussian components from the GMM fit do not reproduce very well the shape of the distribution at $\Delta$(F275W-F438W)$\gtrsim 0.2$ mag. ``Bright'' UV-dim stars here are defined as having a $\Delta$(F275W-F438/5W)$>$0.24 mag (see red dotted line in the right panel of Fig. \ref{fig:uv-dim-1850-bright}), where a detachment from the two main distributions is observed. The bright sample of UV-dim stars in NGC~1850 is indicated with red filled diamonds in the left panel of Fig. \ref{fig:uv-dim-1850-bright}. Finally, we merged the faint and bright UV-dim stars samples of NGC~1850, to obtain the total fraction of UV-dim stars. 

Next, we estimated the fraction of the selected UV-dim stars over the total TO stars for all the clusters in the sample. 
We found that UV-dim stars are $\sim$7.5\% for NGC~1850 ($\sim$ 100 Myr old cluster with eMSTO, e.g. \citealt{bastian16}), $\sim$10.1\% for NGC~1783 ($\sim$ 1.5 Gyr old cluster with eMSTO, e.g. \citealt{milone09}), while this number drops significantly to $\sim$0.5\% for both NGC~1978 and NGC~2121 ($>$ 2 Gyr old and no eMSTO, \citealt{martocchia18b}). 
Very recently, a similar fraction of UV-dim stars for NGC~1783 ($\sim$7\%) has also been reported by \cite{milone22}, confirming our findings.

As the precise fractions might depend on the different selection thresholds, it is important not to stress the absolute fractions, but rather the difference in fractions between the younger and older clusters of the sample (see next Section \ref{sec:disc}). 

Figure \ref{fig:shell_1850} shows the F275W-F438W vs. F814W CMD of NGC~1850 zoomed in on the MSTO region. Our total selected UV-dim stars are represented by red filled circles.  
Additional information from MUSE spectroscopy, based on the work by \cite{kamann22}, is reported on the Figure for our photometrically selected UV-dim stars. Green open diamonds indicate the MUSE spectroscopically identified shell stars, while spectroscopically identified Be stars are indicated as yellow open squares. Finally, black open circles represent the photometric UV-dim stars that have a counterpart in the MUSE spectroscopic sample. The confirmed rapidly rotating, shell stars are shifted to the red also in this CMD including the F275W filter. Hence, we conclude that the UV-dim stars in NGC~1850 are shell stars. Indeed, the majority of our selected UV-dim stars are either shell stars or Be stars (75\% and 85\%, respectively), as shown in Fig. \ref{fig:shell_1850}. Ten stars are identified as UV-dim stars from photometry and are in the MUSE sample, but they are not classified either as a Be star or a shell star. These are represented as the red filled circles with the open black circle in Fig. \ref{fig:shell_1850}. However, these stars are almost all located at the edge of the MSTO (9 out of 10), thus it might be likely that they are only partially covered by the disk in our line of sight. We will discuss more about these results in the next Section \ref{sec:disc}.


\section{Discussion and Conclusions}
\label{sec:disc}

In this paper, we estimated the fraction of UV-dim stars in four massive star clusters in the LMC, namely NGC~1850, NGC~1783, NGC~1978 and NGC~2121, through HST photometry. UV-dim stars are stars that appear fainter in UV colours with respect to the main bulk of the MSTO stars (specifically, we used the F275W-F438/5W\footnote{We also checked other colour-filter combinations, such as F336W-F438W or F343N-F438W, F336W-F814W and we still found that these stars are systematically shifted to the red of the TO.}). In Section \ref{sec:res}, we reported that the fraction of UV-dim stars for NGC~1850 is $\sim$7.5\%, for NGC~1783 it is $\sim$10\%, while both NGC~1978 and NGC~2121 show a negligible fraction of such stars (dropping to $\sim$0.5\%). We find a significant fraction of UV-dim stars in the star clusters that show an eMSTO, namely NGC~1850 and NGC~1783. 
On the contrary, we do not observe UV-dim stars in NGC~1978 and NGC~2121, which do not have an eMSTO. 

\begin{figure*}
\centering
\includegraphics[scale=0.5]{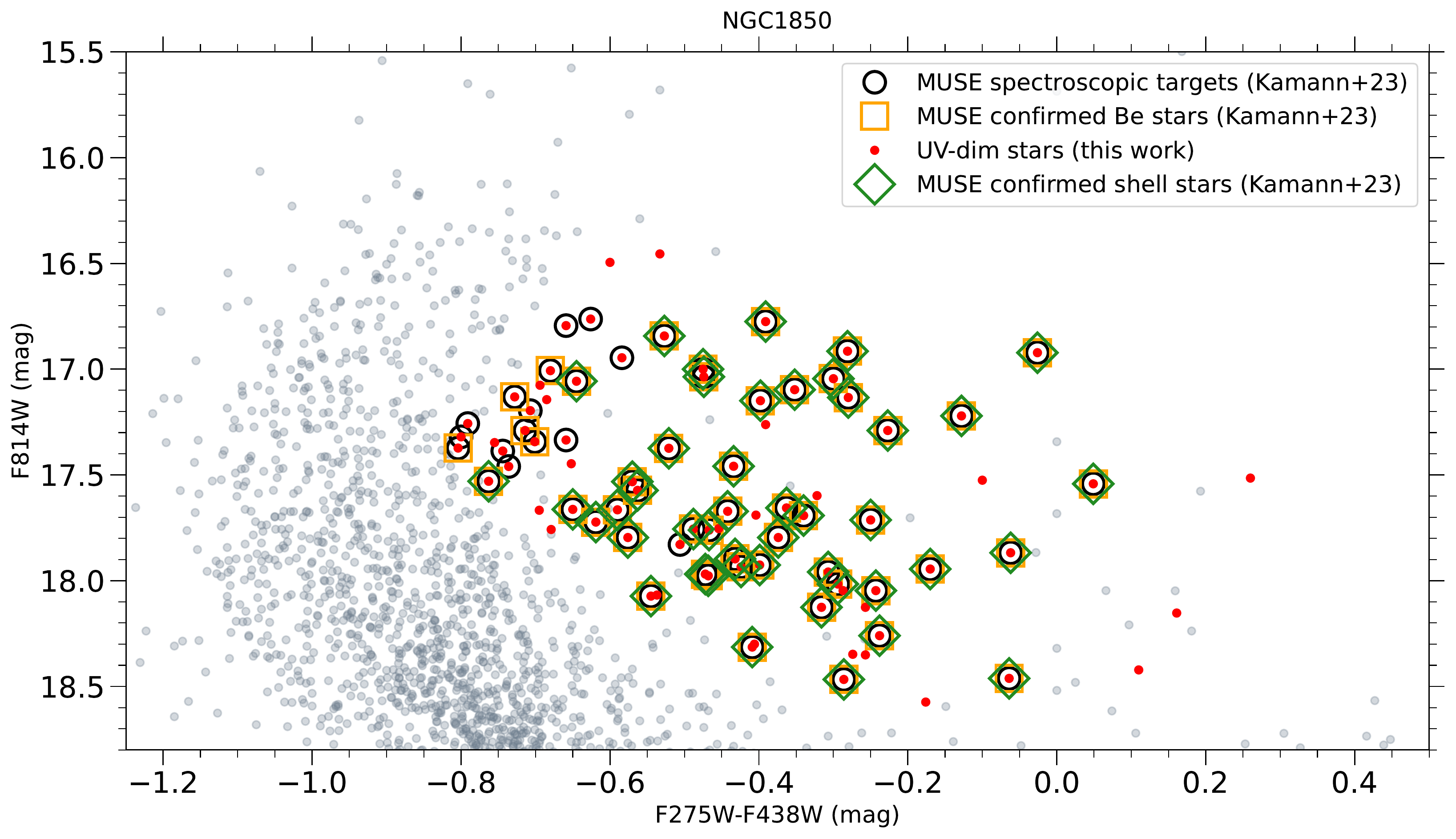}
\caption{F275W-F438W vs F814W CMD of NGC~1850 zoomed in on the MSTO region. Filled red circles represent the identified UV-dim stars based on the photometry selection (this work). Open black circles outline the UV-dim stars that have an available MUSE spectrum. Green (yellow) open diamonds (squares) indicate the identified shell (Be) stars from \protect\cite{kamann22} in our UV-dim stars sample.}
\label{fig:shell_1850}
\end{figure*}

As mentioned in Section \ref{sec:intro}, \cite{kamann22} combined HST photometry and MUSE spectroscopy to measure the rotational velocities of more than 2,000 stars on the MSTO and MS of NGC~1850. 
They found a populous amount of fast rotating Be stars within their sample, out of which 23\% are shell stars, i.e. stars that are seen nearly equator-on and thus are extincted by their own decretion disks.
Their Fig. 10 shows that these stars are shifted to the red in the F336W-F438W vs F438W CMD, hence they appear to be dim in the UV. Additionally, Figure \ref{fig:shell_1850} shows that the majority ($\sim$75\%) of our photometrically selected UV-dim stars, that have a MUSE spectrum available, are shell stars in NGC~1850. Consequently, we concluded that the UV-dim stars selected in NGC~1850 are shell stars. 

It is then straightforward to connect the UV-dim stars to the same rotation phenomenon in the shell stars found in NGC~1850. 
In intermediate age star clusters (like NGC~1783, $\sim$1.5 old Gyr that shows an eMSTO), such fast rotating stars should appear as Ae or Fe stars (with mass $\gtrsim$1.5 \msun), which are also expected to have decretion disks, like those observed in B stars (e.g. \citealt{jaschek86,slettebak82}). However, B stars are able to ionise their disks because of their higher surface temperatures, hence their emission is easier to observe (as Be stars). A and F stars, on the other hand, cannot ionise their disks and thus shell A/F stars are expected to be more difficult to detect. \cite{kamann22} suggested that such stars in intermediate age clusters might be detectable due to infrared excess in their spectral energy distributions. However, as we suggest in this work, a fraction of these stars with decretion disks are expected to be seen nearly equator-on, which will self-extinct the stellar light. Hence, they are expected to be UV-dim, exactly as observed in NGC 1783. 

We then suggest here that the UV-dim stars found in NGC~1850 and NGC~1783 are indeed shell-like stars, rapidly rotating, that are seen nearly equator-on and thus extincted by their own decretion disks. The evidence reported in this paper shows that these stars disappear in clusters older than $\sim$2 Gyr, where fast rotators are expected to disappear. We also note here that, if the UV-dim stars are due to stellar rotation, their appearance/disappearance should depend on stellar mass, i.e. the UV-dim stars should disappear at similar stellar masses (around $\lesssim$1.5 \msun) in all clusters. We observe that UV-dim stars are present around magnitudes F814W$\sim$21 mag in NGC~1783, however they disappear at the same magnitude for NGC~1978 and NGC~2121. This is most likely due to the fact that we are very close to the $\sim$ 1.5 \msun\, limit and the transformation from magnitudes to stellar masses depends on the assumed stellar evolution models and the uncertainty on distance and extinction values.

The results presented here should encourage spectroscopic follow up of such UV-dim photometrically identified stars in NGC~1783, in order to verify whether they are fast rotators, shell-like stars. 
As mentioned above, we would expect that shell features are harder to find in these older clusters, because on the MSTOs such stars have lower temperatures (A and F stars) and they are not able to ionise their disk and behave as strong emitters. In such stars the shell signature is composed of narrow absorption lines that are superimposed on a broad-lined spectrum which represents the photosphere of the star (e.g. \citealt{dominy77, fekel03}). Hence, high signal-to-noise, high resolution spectra are needed to verify whether the UV-dim stars in NGC~1783 are shell stars. However, one could also look at their projected rotational velocities and verify whether such UV-dim stars are rapidly rotating.

\section*{Acknowledgements}
We thank the anonymous referee for a constructive report. SM was supported by a Gliese Fellowship at the Zentrum f\"ur Astronomie, University of Heidelberg, Germany. SK acknowledges funding from UKRI in the form of a Future Leaders Fellowship (grant no. MR/T022868/1). SS acknowledges funding from STFC under the grant no. R276234.

\section*{Data Availability}
All data underlying this article are available on
reasonable request to the corresponding author. The HST photometric catalogue and MUSE spectra of NGC~1850 are publicly available via the CDS portal at \url{https://cdsarc.cds.unistra.fr/viz-bin/cat/J/MNRAS/518/1505}.



\bibliographystyle{mnras}
\bibliography{biblio} 





\bsp	
\label{lastpage}
\end{document}